\documentclass{ptephy_v1}
\usepackage{lineno}
\usepackage[italicdiff]{physics}
\usepackage[colorlinks,citecolor=blue,linkcolor=blue,urlcolor=blue]{hyperref}
\usepackage{bm,amsfonts,siunitx,slashed,mathtools,amssymb,here}
\preprintnumber{YITP-23-140}

\begin{document}
\title{Phase diagram near the quantum critical point in Schwinger model at $\theta = \pi$: \\
analogy with quantum Ising chain}

\author{Hiroki Ohata}
\affil{Yukawa Institute for Theoretical Physics, Kyoto University,\\ 
Sakyo-ku, Kyoto 606-8502, Japan
\email{hiroki.ohata@yukawa.kyoto-u.ac.jp}}

\begin{abstract}
The Schwinger model, one-dimensional quantum electrodynamics, has CP symmetry at $\theta = \pi$ due to the topological nature of the $\theta$ term.
At zero temperature, it is known that as increasing the fermion mass, the system undergoes a second-order phase transition to the CP broken phase, which belongs to the same universality class as the quantum Ising chain.
In this paper, we obtain the phase diagram near the quantum critical point (QCP) in the temperature and fermion mass plane using first-principle Monte Carlo simulations, while avoiding the sign problem by using the lattice formulation of the bosonized Schwinger model.
Specifically, we perform a detailed investigation of the correlation function of the electric field near the QCP and find that its asymptotic behavior can be described by the universal scaling function of the quantum Ising chain.
This finding indicates the existence of three regions near the QCP, each characterized by a specific asymptotic form of the correlation length, and demonstrates that the CP symmetry is restored at any nonzero temperature, entirely analogous to the quantum Ising chain.
The range of the scaling behavior is also examined and found to be particularly wide.
\end{abstract}

\maketitle

\section{Introduction}
Phase transitions are a fundamental concept in physics, observed in numerous systems, including water, magnets, superconductors, and more.
These transitions occur on a variation of an external control parameter, 
such as temperature, magnetic field, pressure, and coupling constant.
Identifying and characterizing the phases in the parameter plane, i.e., obtaining the phase diagram, is a crucial step toward a comprehensive understanding of the system.

The critical point is particularly intriguing in the phase diagram.
Remarkably, in the vicinity of the critical point, various microscopic models sharing the same symmetry pattern and dimensionality exhibit qualitatively the same behaviors, characterized by the same critical exponents, due to the emergent scale invariance at the critical point.
Thus, universality provides a unified understanding of various microscopic models near the critical point and also enables us to analyze critical behaviors of a complex system in terms of a much simpler, and sometimes exactly solvable, model.

In the context of high-energy physics, several model studies suggest the existence of a critical point at finite temperature and baryon chemical potential region in the phase diagram of strongly interacting matter governed by quantum chromodynamics (QCD)~\cite{Asakawa:1989bq, Barducci:1989wi, Barducci:1994, Halasz:1998qr, Berges:1998rc, harada:1998zq, Kitazawa:2002jop}, 
and universality has been applied to the study of critical behaviors~\cite{Pisarski:1983ms, Hatta:2002sj}.
However, although the Monte Carlo simulation based on the lattice formulation of QCD~\cite{Wilson:1974sk, Creutz:1980zw} is at present the only first-principle method to explore the phase structure of QCD,
the conventional lattice QCD simulation suffers from the sign problem at finite baryon chemical potential.
Consequently, the location or even the existence of the QCD critical point has not been established yet.
The detection of the QCD critical point has been one of the main purposes of relativistic heavy-ion collision experiments~\cite{Pandav:2022xxx}.

In this paper, we explore a phase diagram in a prototype gauge theory that possesses a critical point, the Schwinger model~\cite{Schwinger:1962tp}.
The Schwinger model is one-dimensional quantum electrodynamics and shares many low-energy phenomena with QCD, such as confinement, chiral symmetry breaking, and nontrivial topological $\theta$ vacuum.
Due to the topological nature of the $\theta$ term, the Schwinger model has CP symmetry at $\theta = \pi$.
At zero temperature, it is already established that as increasing the fermion mass, the system undergoes a second-order phase transition to the CP broken phase, which belongs to the same universality class as the quantum Ising chain~\cite{Hamer:1982mx, Byrnes:2002nv, Shimizu:2014fsa, Azcoiti:2017mxl, Thompson:2021eze}.
However, compared to the comprehensive studies on the critical point itself, the phase structure near the critical point at finite temperature remains much less known.

Our present analysis is based on universality with the quantum Ising chain and first-principle Monte Carlo simulations.
Although the Monte Carlo simulation of the Schwinger model at $\theta = \pi$ is hindered by the severe sign problem when using the conventional lattice formulations,
the author recently pointed out~\cite{Ohata:2023sqc} that this problem can be avoided by using the lattice formulation of the bosonized Schwinger model~\cite{Bender:1984qg}.
In this work, we obtain the phase diagram of the Schwinger model near the critical point by verifying the scaling behavior of the correlation function, which is predicted from universality with the quantum Ising chain, through Monte Carlo simulations.
Our study provides a comprehensive understanding of the phase structure near the critical point in the Schwinger model at $\theta = \pi$ and demonstrates the power of universality in a gauge theory.

This paper is organized as follows.
In section~\ref{sec:spontaneous_CP_breaking}, we review the spontaneous CP symmetry breaking at zero temperature from both analytical and numerical sides, along with introducing the Schwinger model and the $\theta$ term.
In section~\ref{sec:phase_diagram}, we first review the quantum Ising chain and then conjecture the phase diagram of the Schwinger model at $\theta = \pi$ based on universality with the quantum Ising chain.
Our strategy to establish the conjectured phase diagram using the Monte Carlo method is also explained in this section.
In section~\ref{sec:numerical}, we perform a detailed numerical investigation of the correlation function and establish the conjectured phase diagram.
Section~\ref{sec:summary} is devoted to summary and conclusion.

\section{Spontaneous CP symmetry breaking at zero temperature} \label{sec:spontaneous_CP_breaking}
In this section, we briefly review the spontaneous CP symmetry breaking at zero temperature in the Schwinger model at $\theta = \pi$.

Let us begin by introducing the Schwinger model and the $\theta$ term.
The Schwinger model is a one-dimensional $\mathrm{U}(1)$ gauge theory coupled with a Dirac fermion~\cite{Schwinger:1962tp}.
The Euclidean action of the Schwinger model is given by
\begin{equation}
S_E = \int d^2 x\, \overline{\psi}\qty(\slashed{\partial} + g \slashed{A} + m)\psi + \frac{1}{4} F_{\mu\nu} F_{\mu\nu}. \label{eq:Schwinger}
\end{equation}
Here $F_{\mu\nu} = \partial_\mu A_\nu - \partial_\nu A_\mu$ is the field strength of the $\mathrm{U}(1)$ gauge field $A_\mu$, $\psi$ the Dirac fermion, $g$ the dimensionful gauge coupling, and $m$ the fermion mass.
The Euclidean action~(\ref{eq:Schwinger}) is symmetric under the CP transformation.
It is possible to introduce the so-called $\theta$ term without breaking the $\mathrm{U}(1)$ gauge symmetry:
\begin{equation}
i \theta Q, \quad Q = \int d^2x \, \frac{g}{4\pi} \epsilon_{\mu\nu} F_{\mu\nu},
\end{equation}
where $\epsilon_{\mu\nu}$ is the antisymmetric tensor.
Introducing the $\theta$ term into the action typically results in the explicit breaking of the CP symmetry, 
since the term is antisymmetric under the CP transformation.
However, there is one nontrivial exception: the CP symmetry is not explicitly broken at $\theta = \pi$ because the $\theta$ term appears as $\exp(i\theta Q)$ in the path-integral, and $Q$ is known to take an integer value for any gauge configuration.
The fate of this CP symmetry is one of the key issues in this paper.

\begin{figure}[htb]
\centering
\includegraphics[width = 0.45\textwidth]{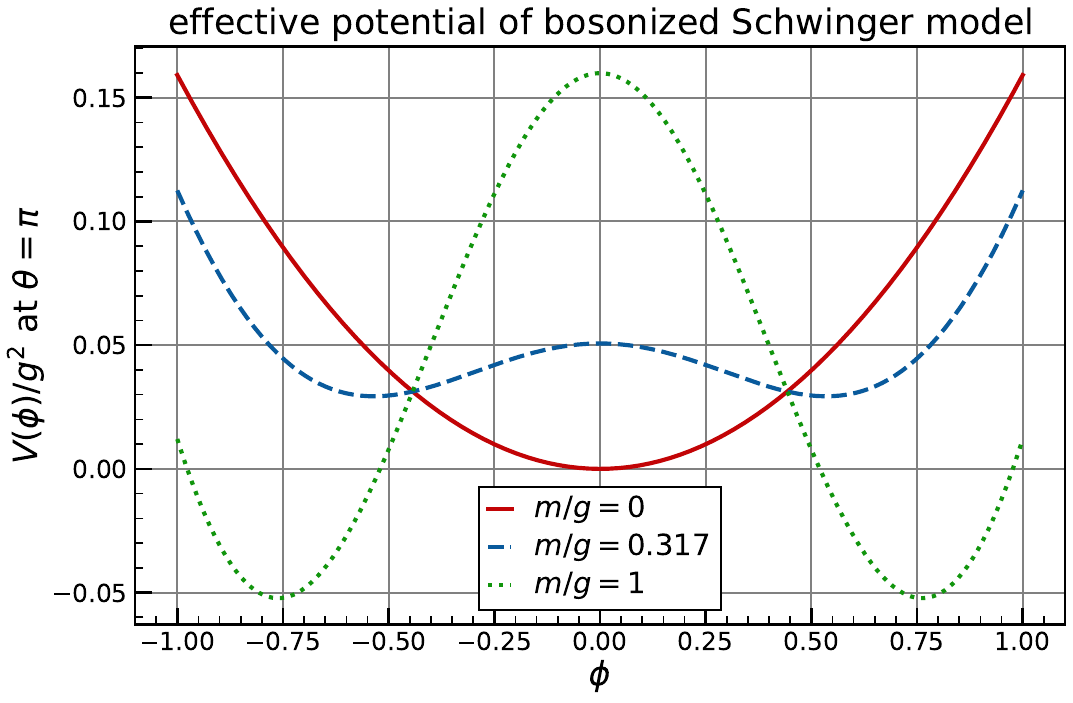}

\caption{
The effective potential of the bosonized Schwinger model~(\ref{eq:effective_potential}) at $\theta = \pi$.
\label{fig:effective_potential}
}
\end{figure}

The spontaneous CP symmetry breaking at $\theta = \pi$ was first predicted by Coleman with an intuitive analytical picture using the bosonized form of the Schwinger model~\cite{Coleman:1976uz}.
After bosonization and integrating out the gauge fields, the Hamiltonian of the bosonized Schwinger model reads~\cite{Coleman:1974bu, Mandelstam:1975hb, Coleman:1975pw}
\begin{equation}
H = \int d x\, \frac{1}{2}\pi^2 + \frac{1}{2} \qty(\partial_x \phi)^2 + \frac{g^2}{2\pi} \phi^2
- \frac{e^\gamma}{2 \pi^{3/2}} m g \mathcal{N}_{g / \sqrt{\pi}} \cos(2\sqrt{\pi} \phi - \theta). \label{eq:continuum_hamiltonian}
\end{equation}
Here $\pi$ is the conjugate momentum, $\gamma$ Euler's constant,
and $\mathcal{N}_{g / \sqrt{\pi}}$ denotes the normal ordering with respect to the boson mass $g / \sqrt{\pi}$.
In this formulation, the scalar field $\phi$ is related to the electric field $E = F_{01}$ as
\begin{equation}
\frac{\phi}{\sqrt{\pi}} = \frac{E}{g}, \label{eq:scalar}
\end{equation}
through the Gauss law constraint. 
\footnote{The fermion density operator is expressed by the scalar field as $\psi^\dag \psi = \partial_x \phi / \sqrt{\pi}$~\cite{Coleman:1974bu, Mandelstam:1975hb}.
By solving the Gauss law $\partial_x E / g = \psi^\dag \psi$, we obtain Eq.~(\ref{eq:scalar}).
The constant of integration is interpreted as the external electric field and absorbed into $\theta$~\cite{Coleman:1976uz}.}
Thus, $\phi$ serves as an order parameter of the CP symmetry.
The effective potential of $\phi$ reads~\cite{Coleman:1976uz}
\begin{equation}
V(\phi) = \frac{g^2}{2\pi} \phi^2 - \frac{e^\gamma}{2\pi^{3/2}} m g \cos(2\sqrt{\pi} \phi - \theta). \label{eq:effective_potential}
\end{equation}
At $\theta = 0$, the cosine term forms a pocket at $\phi = 0$, indicating a unique CP-symmetric vacuum at any fermion mass.
This changes drastically at $\theta = \pi$.
Figure~\ref{fig:effective_potential} shows the effective potential at $\theta = \pi$ at various fermion masses.
For small fermion masses, the bosonic mass term dominates the potential, and we expect a unique vacuum.
On the other hand, for large fermion masses, the cosine term forms a potential wall at $\phi = 0$,  
and we expect degenerate vacua, corresponding to the spontaneous CP symmetry breaking.
In an intermediate fermion mass region where the CP symmetry is either maintained or at least well-preserved, 
by expanding the cosine to the second order,
the effective potential becomes~\cite{Byrnes:2002nv}
\begin{equation}
V(\phi) \simeq \frac{1}{2} \qty{\frac{g}{\sqrt{\pi}} \qty(1 - \sqrt{\pi} e^\gamma \frac{m}{g})}^2 \phi^2.
\end{equation}
As $m / g$ increases, the effective mass decreases and reaches zero at  
\begin{equation}
\frac{m_c}{g} = \frac{1}{\sqrt{\pi}e^\gamma} \simeq 0.317, \label{eq:analytic_critical}
\end{equation} 
where we expect a second-order phase transition.
It is important to note that this second-order phase transition is driven purely by quantum fluctuations.
Hence, the critical point at zero temperature $(m_c, T = 0)$ is referred to as the quantum critical point (QCP) in the following.

First-principle numerical simulations are needed for a more quantitative and conclusive understanding.
Unfortunately, the conventional Monte Carlo method is not applicable due to the severe sign problem at $\theta = \pi$, 
when using the Euclidean lattice fermion formulations.
Consequently, numerical investigations of the QCP have been mainly performed using the finite-dimensional spin Hamiltonian, 
which is derived by applying the Jordan--Wigner transformation to the Kogut--Susskind Hamiltonian formulation of the Schwinger model~\cite{Kogut:1974ag, Hamer:1982mx, Ranft:1982bi, Byrnes:2002nv, Buyens:2017crb}.

The first numerical evidence of the above analytical picture was provided by Hamer et al.~\cite{Hamer:1982mx}, in which the spin Hamiltonian was combined with the finite-size scaling method~\cite{Hamer:1980} to identify the QCP.
They estimated the critical mass to be $m_c / g = 0.325(20)$ and a critical exponent $\nu = 0.9(1)$.
A more precise investigation was done by Byrnes et al.~\cite{Byrnes:2002nv} using the density matrix renormalization group~\cite{White:1992zz, White:1993zza}.
They obtained at present the most precise estimate of the critical mass 
\begin{equation}
\frac{m_c}{g} = 0.3335(2), \label{eq:critical_mass}
\end{equation}
and two critical exponents
\begin{equation}
\nu = 1.01(1), \quad \frac{\beta}{\nu} = 0.125(5).
\end{equation}
These two critical exponents precisely agree with those of the quantum Ising chain ($\nu = 1, \beta = 1/8$), indicating that these two models are in the same universality class.
As a quite different approach, Shimizu and Kuramashi~\cite{Shimizu:2014fsa} applied the Grassmann tensor renormalization group~\cite{Gu:2010gr} to the conventional Wilson fermion formulation in Euclidean space-time.
Using the Lee--Yang and Fisher zero analyses, they provided further evidence that the QCP belongs to the Ising universality class.
Other approaches include, for example, a conversion method from the imaginary $\theta$ term~\cite{Azcoiti:2017mxl}, and one using quantum computing~\cite{Thompson:2021eze}.

\section{Phase diagram near the quantum critical point} \label{sec:phase_diagram}
Compared to the comprehensive studies on the second-order phase transition at zero temperature, 
the phase structure near the QCP remains much less known.
To the best of the author's knowledge, 
the only study addressing this issue was conducted by Buyens et al.~\cite{Buyens:2016ecr}, 
in which they investigated the fate of the spontaneously broken CP symmetry at finite temperature using the matrix product operators~\cite{Verstraete:2004ve, Zwolak:2004zw} based on the spin Hamiltonian formulation.
Through a detailed investigation of the electric field at finite temperature near $\theta = \pi$, they obtained an indication that the CP symmetry is restored at any nonzero temperature.

In this paper, we explore broad characteristics near the QCP from the perspective of universality with the quantum Ising chain using a completely different lattice formulation.
In this section, we first review the quantum Ising chain, in particular its phase structure at finite temperature, and then explain our strategy to explore the phase diagram of the Schwinger model at $\theta = \pi$.

\subsection{Quantum Ising chain} \label{subsec:Ising}
We here review the phase diagram of the quantum Ising chain following a review~\cite{Sachdev:2011}.
The quantum Ising chain is the quantum analog of the two-dimensional classical Ising model with no external magnetic field. 
The Hamiltonian of the quantum Ising chain is given by
\begin{equation}
H_I = -J \sum_{i_x} \qty( \sigma_{i_x}^z \sigma_{i_x + 1}^z + g \sigma_{i_x}^x),
\end{equation}
where $J > 0$ is an overall energy scale, $g > 0$ is the dimensionless coupling constant, and $\sigma_{i_x}^{z}, \sigma_{i_x}^{x}$ are the Pauli matrixes
\begin{equation}
\sigma_{i_x}^z = \mqty(\dmat[0]{1, -1}), \quad \sigma_{i_x}^x = \mqty(\admat[0]{1, 1}),
\end{equation}
which reside on a spatial site $i_x$.
The Hamiltonian is symmetric under the $\mathbb{Z}_2$ transformation, generated by the unitary operator $\prod_{i_x} \sigma_{i_x}^x$.

Unlike the massive Schwinger model, the quantum Ising chain is exactly solvable in the sense that all eigenstates are obtained analytically by applying the Jordan--Wigner transformation and diagonalizing the Hamiltonian by the Bogoliubov transformation~\cite{Lieb:1961fr, Pfeuty:1970pf}.
At zero temperature, as decreasing the coupling constant, the system undergoes a second-order phase transition to the ferromagnetic ($\mathbb{Z}_2$ broken) phase at $g_c = 1$, which is analogous to the Schwinger model at $\theta = \pi$.

\begin{figure}[htb]
\centering
\includegraphics[width = 0.45\textwidth]{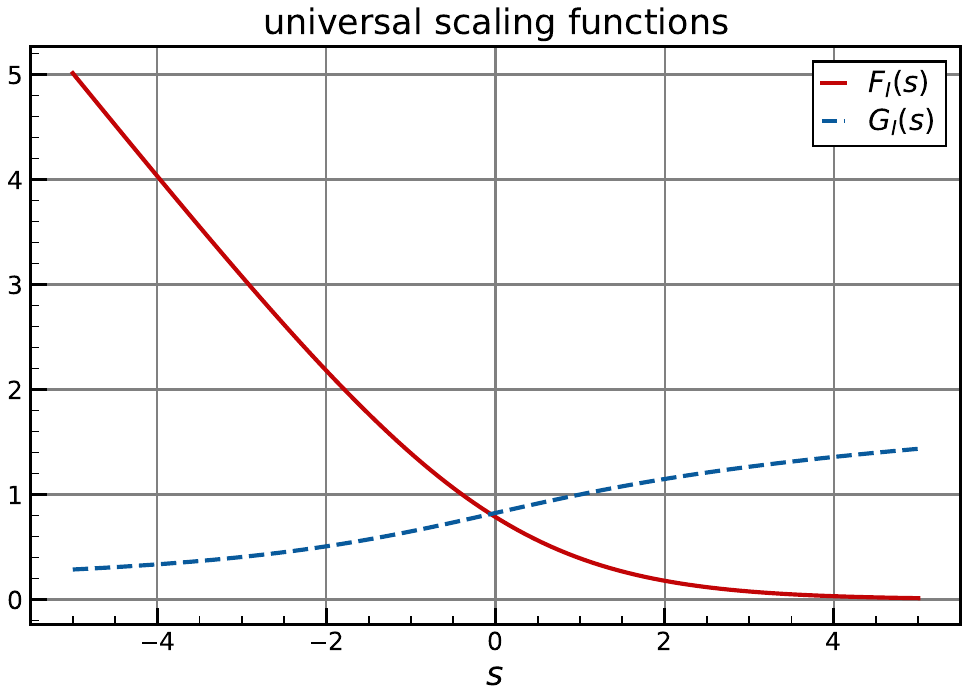}

\caption{
The universal scaling functions of the quantum Ising chain~(\ref{eq:F}, \ref{eq:G}).
\label{fig:universal_analytic}
}
\end{figure}

The phase diagram of the quantum Ising chain in the temperature and coupling constant plane can be deduced from the correlation function of the order parameter at finite temperature
\footnote{Note that we do not subtract the connected part $\expval{\sigma^z}^2$ from the correlation function.}
\begin{equation}
C(x) = \expval{\sigma_0^z \sigma_{i_x}^z} = \left. \tr(\sigma_0^z \sigma_{i_x}^z e^{-H_I / T}) \middle/ \tr(e^{-H_I / T}) \right.. \label{eq:correlation}
\end{equation}
A crucial feature that we will exploit throughout this paper is the explicit asymptotic form of the correlation function, derived by Sachdev~\cite{Sachdev:1995bf}:
\begin{equation}
\lim_{x \to \infty} C(x) = Z T^{1/4} G_I(\Delta / T) \exp(-\frac{Tx}{c} F_I(\Delta / T)), \quad \Delta = r (g_c - g), \label{eq:asymptotic_Ising}
\end{equation}
where $Z, c, r$ are the nonuniversal constants, and $F_I$ and $G_I$ are the universal scaling functions of the quantum Ising chain. 
Their explicit forms are~\cite{Sachdev:1995bf}
\begin{align}
F_I(s) &= \abs{s} \Theta(-s) + \frac{1}{\pi} \int_0^\infty dy\, \ln\coth \frac{\qty(y^2 + s^2)^{1/2}}{2}, \label{eq:F} \\
\ln G_I(s) &= \int_s^1 \frac{dy}{y} \qty[\qty(\frac{d F_I(y)}{dy})^2 - \frac{1}{4}] + \int_1^\infty \frac{dy}{y} \qty(\frac{d F_I(y)}{dy})^2, \label{eq:G}
\end{align}
where $\Theta(x)$ is the step function.
Figure~\ref{fig:universal_analytic} plots the universal scaling functions $F_I, G_I$,
illustrating their smoothness throughout the entire region.

The correlation length is defined from the exponential decay of the correlation function~(\ref{eq:asymptotic_Ising}) as
\begin{equation}
\xi^{-1} = \frac{T}{c} F_I\qty(\frac{\Delta}{T}).
\end{equation}
At any nonzero temperature, the correlation length is finite, 
indicating that the correlation function vanishes in the long-distance limit.
Simultaneously, the correlation function should converge to its connected part as
\begin{equation}
\lim_{x \to \infty} C(x) = \expval{\sigma^z}^2.
\end{equation}
Thus, we can deduce that the system is in the paramagnetic ($\mathbb{Z}_2$ symmetric) phase at any nonzero temperature.

\begin{figure}[htb]
\centering
\includegraphics[width = 0.45\textwidth]{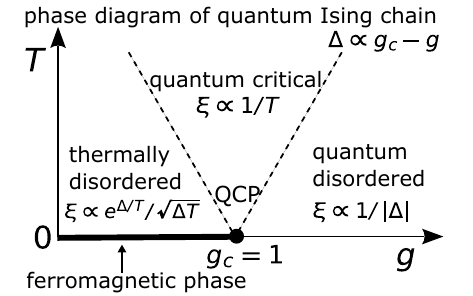}
\quad
\includegraphics[width = 0.45\textwidth]{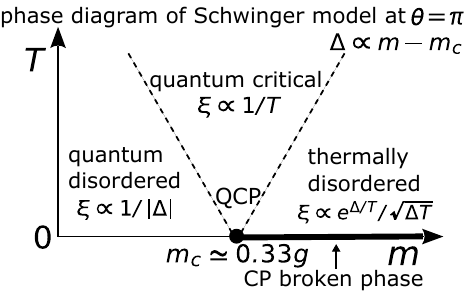}

\caption{
(Left) Phase diagram of the quantum Ising chain. (Right) Conjectured phase diagram of the Schwinger model at $\theta = \pi$.
\label{fig:phase_diagram}
}
\end{figure}

In the vicinity of the QCP, even a slight shift in the coupling constant results in a substantial change in $\Delta / T$.
It is meaningful to categorize the phase into three depending on the value of $\Delta / T$, 
each characterized by a specific asymptotic form of the correlation length~\cite{Sachdev:1995bf}:
\begin{equation}
\xi = \begin{cases}
c\sqrt{\frac{\pi}{2\Delta T}} e^{\Delta / T}, & \Delta / T \gg 1, \\
\frac{4c}{\pi T}, & \abs{\Delta} / T \ll 1, \\
\frac{c}{\abs{\Delta}}, & \Delta / T \ll -1.
\end{cases} \label{eq:three_length}
\end{equation}
In the low-temperature region with $\Delta > 0$, the correlation length diverges exponentially toward the zero temperature limit, corresponding to the ferromagnetic phase at zero temperature.
Conversely, the long-range order at zero temperature is thermally destroyed.
This region is commonly referred to as the thermally disordered region in the context of condensed matter physics~\cite{Vojta:2003}.
In the other low-temperature region with $\Delta < 0$, the correlation length saturates at a finite value in the zero temperature limit, resulting in a disordered phase even at zero temperature.
This region is called the quantum disordered phase~\cite{Vojta:2003}.
The final region with $\abs{\Delta} / T \ll 1$ is the so-called quantum critical region, where the physics is considered to be governed by the thermal excitations of the quantum critical ground state~\cite{Vojta:2003}.
The resulting phase diagram of the quantum Ising chain is shown in the left panel of Fig.~\ref{fig:phase_diagram}.
The universality with the quantum Ising chain implies the phase diagram of the Schwinger model at $\theta = \pi$ near the QCP, as shown in the right panel of Fig.~\ref{fig:phase_diagram}.

\subsection{Schwinger model at $\theta = \pi$} \label{subsec:Schwinger}
We here explain our strategy to establish the conjectured phase diagram of the Schwinger model at $\theta = \pi$.
In the previous subsection, we observed that the phase diagram of the quantum Ising chain can be deduced from the asymptotic form of the correlation function~(\ref{eq:asymptotic_Ising}).
Therefore, our approach is rather straightforward: we calculate the correlation function of the electric field at $\theta = \pi$ at finite temperatures near the QCP and examine whether the correlation function shares the same asymptotic form as the quantum Ising chain~(\ref{eq:asymptotic_Ising}).
If this confirmation is obtained at a certain temperature, the scaling behavior should also hold at lower temperatures, allowing us to determine the phase diagram of the Schwinger model near the QCP.
Our method does not aim to approach the QCP itself, but rather inspect the scaling behavior near the QCP.
Therefore, we need not employ the finite-size scaling method because the correlation length is always finite in our analysis.
The most significant difficulty is that we must circumvent the sign problem at $\theta = \pi$ in some way.

Recently, the author pointed out~\cite{Ohata:2023sqc} that by utilizing the lattice formulation of the bosonized Schwinger model~\cite{Bender:1984qg}, 
Monte Carlo simulations can be performed without encountering the sign problem.
The lattice bosonized Schwinger model is obtained after removing the normal ordering appearing in the cosine term in the bosonized Hamiltonian~(\ref{eq:continuum_hamiltonian}).
In the continuum, using the Wick theorem, the normal ordering can be removed as~\cite{Coleman:1974bu}
\begin{equation}
\mathcal{N}_{g / \sqrt{\pi}} \exp(i 2 \sqrt{\pi} \phi) = \exp{2\pi \Delta\qty(x = 0; \frac{g}{\sqrt{\pi}} )} \exp(i 2 \sqrt{\pi} \phi),
\end{equation}
where $\Delta(x; g / \sqrt{\pi})$ is the Feynman propagator for the scalar field of mass $g / \sqrt{\pi}$.
The lattice counterpart of this relation is obtained by just replacing the Feynman propagator with that on the lattice~\cite{Bender:1984qg}
\begin{equation}
\Delta\qty(x; \frac{g}{\sqrt{\pi}}; \frac{1}{a}) = \int^{\pi}_{-\pi} \frac{d^2 k}{\qty(2 \pi)^2} e^{i k x} \qty(4 \sum_{\mu} \sin^2\qty(\frac{k_\mu}{2}) + \qty(\frac{ag}{\sqrt{\pi}})^2)^{-1},
\end{equation}
where $a$ is the lattice spacing.
Consequently, the lattice counterpart of the continuum Hamiltonian~(\ref{eq:continuum_hamiltonian}) can be expressed without using the normal ordering 
\begin{equation}
H a = \sum_{x = 0}^{L_x - 1} \frac{1}{2} \qty(a \pi_{x})^2 + \frac{1}{2} \qty( \partial_x \phi_{x} )^2 + 
\frac{\qty(ag)^2}{2\pi} \phi_{x}^2
- \frac{e^{\gamma}}{2 \pi^{3/2}} \frac{m}{g} \qty(ag)^2 \mathcal{O}(1/ag) \cos(2 \sqrt{\pi} \phi_{x} - \theta).
\end{equation}
Here, $\partial_x$ denotes the forward derivative $\partial_x f_x \coloneqq f_{x + 1} - f_x$, and the factor $\mathcal{O}(1 / ag)$ is defined as
\begin{equation}
\mathcal{O}(1 / ag) \coloneqq \exp{2 \pi \Delta\qty(0; \frac{g}{\sqrt{\pi}}; \frac{1}{a})}, \\
\end{equation}
and behaves $\mathcal{O}(1/ag) \simeq 10 / ag$ for $ag \ll 1$~\cite{Ohata:2023sqc}.
The validity of this lattice formulation was explicitly confirmed in Ref.~\cite{Ohata:2023sqc}.

The correlation function of the electric field at temperature $T = 1 / (a L_\tau)$ can be expressed by the path-integral, along with using Eq.~(\ref{eq:scalar}):
\begin{subequations}
\begin{align}
C(x) &= \frac{1}{g^2}\expval{E_{x} E_0} \\
&= \frac{1}{\pi} \expval{\phi_{x} \phi_{0}} \\
&= \frac{1}{\pi} \left. \tr(\phi_x \phi_0 e^{-H / T}) \middle/ \tr e^{-H / T} \right. \\
&= \frac{1}{\pi} \left. \int D \phi \, \phi_{x, 0} \phi_{0, 0} e^{- S_{E}} \middle/ \int D \phi \, e^{-S_{E}}, \right. \label{eq:thermalexpvalue}
\end{align}
\end{subequations}
where $S_E$ is the lattice Euclidean action of the bosonized Schwinger model
\begin{multline}
S_E = \sum_{\tau = 0}^{L_\tau - 1} \sum_{x = 0}^{L_x - 1} \frac{1}{2} \qty( \partial_\tau \phi_{x, \tau})^2 + \frac{1}{2} \qty( \partial_x \phi_{x, \tau} )^2 +  \frac{\qty(ag)^2}{2\pi} \phi_{x, \tau}^2 \\
- \frac{e^{\gamma}}{2 \pi^{3/2}} \frac{m}{g} \qty(ag)^2 \mathcal{O}(1/ag) \cos(2 \sqrt{\pi} \phi_{x, \tau} - \theta). \label{eq:action}
\end{multline}
The periodic boundary condition is imposed for both directions in this paper.
The lattice Euclidean action~(\ref{eq:action}) is obviously real and bounded below even at $\theta \neq 0$, meaning no sign problem.
We generate the Monte Carlo configurations by using the heat-bath algorithm and the rejection sampling in this paper.

\section{Numerical results} \label{sec:numerical}
In this work, we use a sufficiently fine and large lattice of $ag = 0.2, L_x = 1792$.
We later verify that the finite lattice spacing and finite spatial length effects are indeed nearly negligible in our analysis.
As for the temperature, we mainly consider two low temperatures $L_\tau = 112, 56$ to inspect the scaling behavior near the QCP.

\subsection{Autocorrelation and lattice artifacts}
In this subsection, we investigate the autocorrelation and lattice artifacts in preparation for the large-scale numerical simulations in the next subsection.

\begin{figure}[htb]
\centering
\includegraphics[width = 0.45\textwidth]{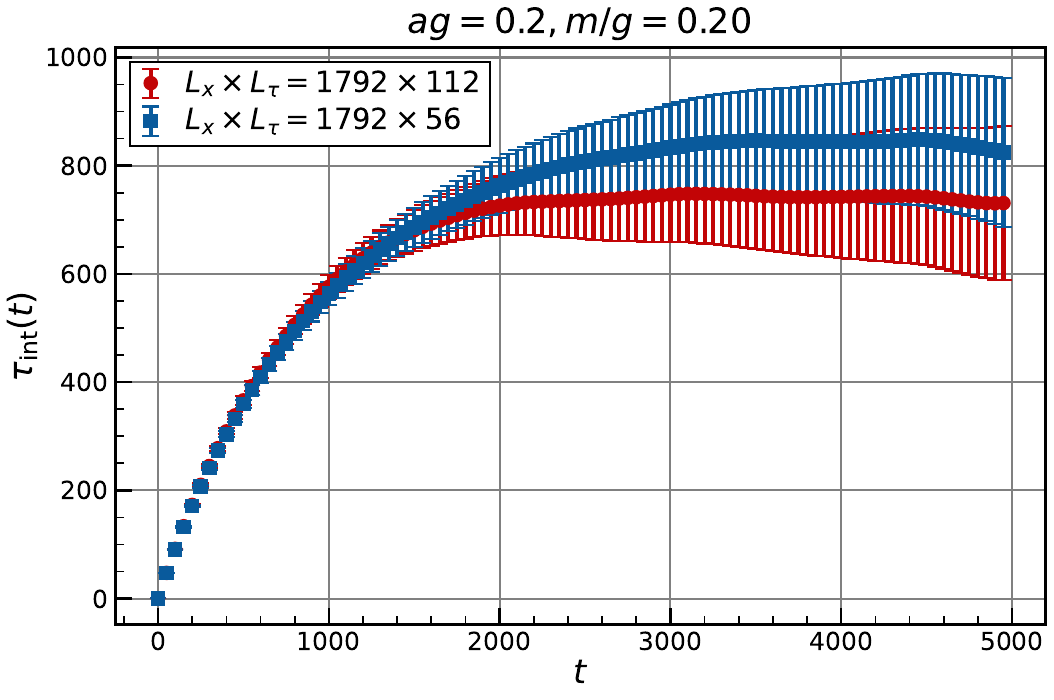}
\qquad
\includegraphics[width = 0.45\textwidth]{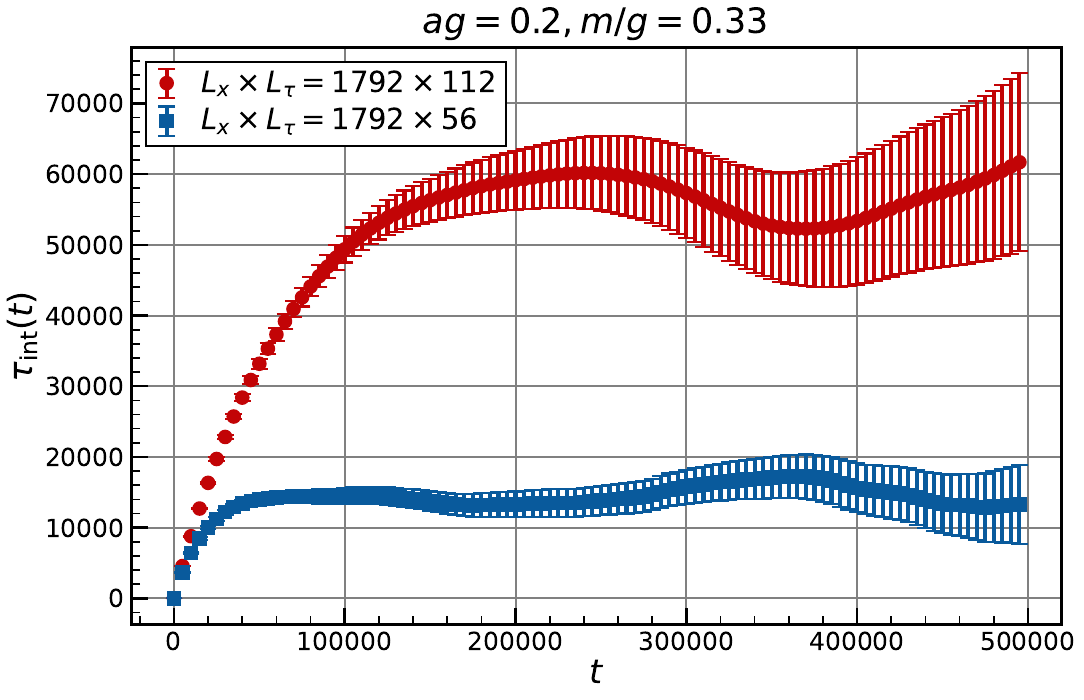}
\qquad
\includegraphics[width = 0.45\textwidth]{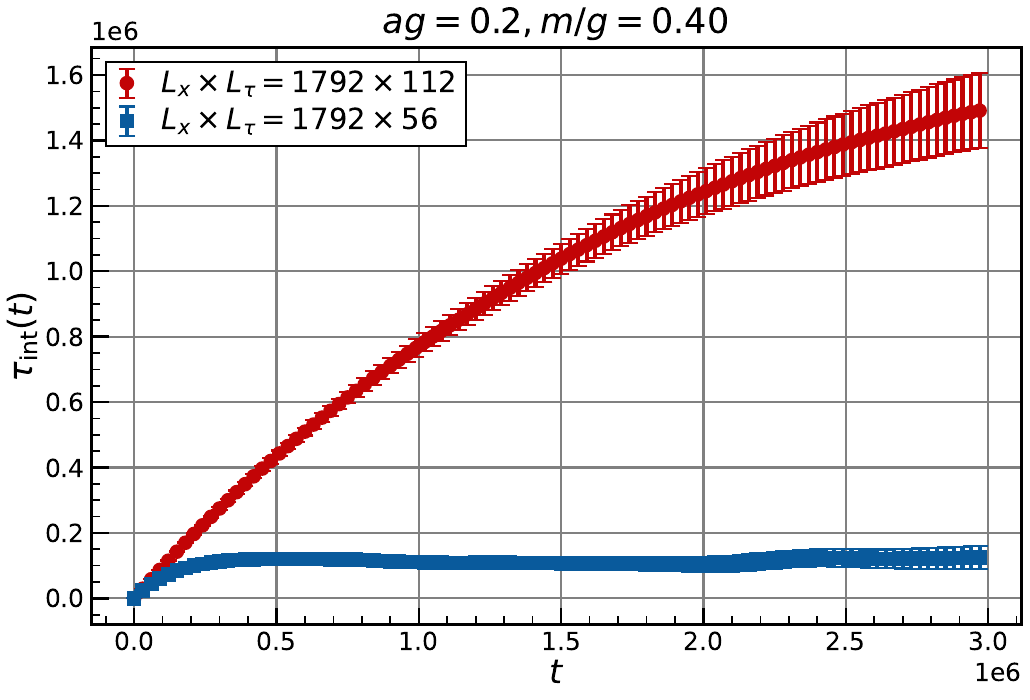}

\caption{
The integrated autocorrelation time with finite lattice data~(\ref{eq:auto_correlation}) at $m / g = 0.2, 0.33, 0.4$.
\label{fig:auto_correlation}
}
\end{figure}

We first investigate the autocorrelation among Monte Carlo configurations near the QCP for reliable error estimates.
The autocorrelation can be characterized by the autocorrelation function~\cite{Sokal:1997}
\begin{equation}
A(t) = \frac{1}{N - t} \sum_{i = 1}^{N - t} \phi_i \phi_{i + t} - \expval{\phi}^2,
\end{equation}
where $\phi_i$ is the mean of $\phi_{x, \tau}$:
\begin{equation}
\phi_i = \frac{1}{L_x L_\tau} \sum_{x, \tau} \phi_{x, \tau}
\end{equation}
at the $i$-th Monte Carlo configuration,
and $\expval{\phi}$ is the subtraction term, which should be $0$ in our analysis.
From the normalized autocorrelation function
\begin{equation}
C(t) = \frac{A(t)}{A(0)},
\end{equation}
the integrated autocorrelation time with finite lattice data is defined as
\begin{equation}
\tau_{\mathrm{int}}(t) = \frac{1}{2} + \sum_{i = 1}^t C(i). \label{eq:auto_correlation}
\end{equation}
Twice the integrated autocorrelation time
\begin{equation}
\tau_{\mathrm{int}} = \lim_{t \to \infty} \tau_{\mathrm{int}}(t) \label{eq:auto_correlation2}
\end{equation}
gives an estimate of the number of iterations required to generate mostly independent configurations.
Figure~\ref{fig:auto_correlation} shows the integrated autocorrelation time~(\ref{eq:auto_correlation}) at $m / g = 0.2, 0.33, 0.4$.
From such behaviors, we estimate the integrated autocorrelation time~(\ref{eq:auto_correlation2}).
The estimated integrated autocorrelation times at the two lattices and various fermion masses are summarized in Table~\ref{table:auto_correlation}.
\begin{table}[!h]
\centering
\begin{tabular}{cccc}
\hline
$ag$ & $L_x \times L_{\tau}$ & $m / g $ & $\tau_{\mathrm{int}}$ \\
\hline
0.2 & $1792 \times 112$ & 0.2  & $\sim 8   \times 10^2$ \\
0.2 & $1792 \times 112$ & 0.33 & $\sim 6   \times 10^4$ \\
0.2 & $1792 \times 112$ & 0.4  & $\sim 2 \times 10^6$ \\
\hline
0.2 & $1792 \times 56$  & 0.1  & $\sim 2   \times 10^2$ \\
0.2 & $1792 \times 56$  & 0.2  & $\sim 8   \times 10^2$ \\
0.2 & $1792 \times 56$  & 0.33 & $\sim 1.4 \times 10^4$ \\
0.2 & $1792 \times 56$  & 0.4  & $\sim 1.2 \times 10^5$ \\
0.2 & $1792 \times 56$  & 0.5  & $\sim 5 \times 10^5$ \\
\hline
\end{tabular}
\caption{Estimated integrated autocorrelation times at the two lattices and various fermion masses.}
\label{table:auto_correlation}
\end{table}
We find that the integrated autocorrelation time is longer at larger fermion masses and lower temperatures, which seems to be related to the expected correlation length near the QCP, as shown in the right panel of Fig.~\ref{fig:phase_diagram}.
In this work, we set the number of iterations to approximately $\tau_{\mathrm{int}} / 2$ and use the binning of ten to eliminate the remaining autocorrelation.
We note that the long autocorrelation time at large fermion mass and low temperature is the main source of our computational costs.
To mitigate this problem, different methods, such as the exchange Monte Carlo, may be useful.

\begin{figure}[htb]
\centering
\includegraphics[width = 0.45\textwidth]{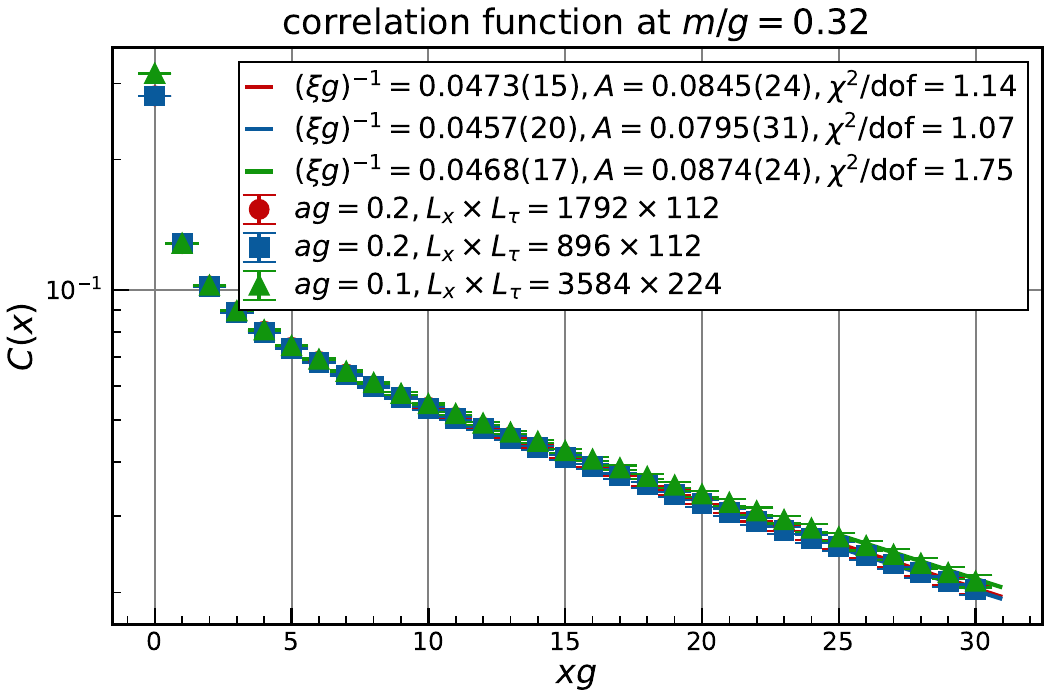}
\caption{
Correlation functions and fit results using the single exponential function~(\ref{eq:exponential}) at three lattices of $ag = 0.2, L_x \times L_\tau = 1792 \times 112$, $ag = 0.2, L_x \times L_\tau = 896 \times 112$, and $ag = 0.1, L_x \times L_\tau = 3584 \times 224$.
For visibility, only one-fifth and one-tenth of the data points are plotted for $ag = 0.2$ and $ag = 0.1$, respectively.
\label{fig:artifact}
}
\end{figure}

We also investigate the finite lattice spacing and finite spatial length effects.
For this purpose, we compare correlation functions at $m / g = 0.32$ calculated using three different lattices:
\begin{itemize}
\item $ag = 0.2, L_x \times L_\tau = 1792 \times 112$,
\item $ag = 0.2, L_x \times L_\tau = 896 \times 112$,
\item $ag = 0.1, L_x \times L_\tau = 3584 \times 224$.
\end{itemize}
By comparing the first and the second lattices,
we can assess the finite spatial length effect.
The finite lattice spacing effect can be examined by comparing the first and the third lattices.
Figure~\ref{fig:artifact} shows the correlation functions at these three lattices.
For each correlation function, we perform a correlated fit using the single exponential function
\begin{equation}
C(x) = A \exp(-\frac{x}{\xi}), \label{eq:exponential}
\end{equation}
with the correlation length $\xi$ and the amplitude $A$ being the fit parameters.
The fit range is set as $xg \in \qty[25, 30]$.
The fit results are also shown in Fig~\ref{fig:artifact}.
We find that the resulting correlation lengths and amplitudes are consistent with each other,
suggesting that both finite lattice spacing and finite spatial length effects are nearly negligible in our analysis.

\subsection{Correlation function near the quantum critical point}
We now perform a detailed investigation of the correlation function near the QCP.
To carefully examine the behavior near the QCP, we analyze correlation functions over a broad range of fermion masses.
The simulation parameters relevant in this subsection are summarized in Table~\ref{table:simulation_parameter}.
\begin{table}[!h]
\centering
\begin{tabular}{ccccc}
\hline
$ag$ & $L_x \times L_{\tau}$ & $m / g $ & num. of iterations & num. of configurations \\
\hline
0.2 & $1792 \times 112$ & 0.16, 0.18, 0.2, 0.22, 0.24, 0.26 & $4 \times 10^2$ & $10^5$ \\
0.2 & $1792 \times 112$ & 0.28, 0.3, 0.32, 0.34, 0.36, 0.38 & $3 \times 10^4$ & $2 \times 10^3$ \\
0.2 & $1792 \times 112$ & 0.4, 0.42, 0.44                   & $10^6$ & $10^3$ \\
\hline
0.2 & $1792 \times 56$  & 0.08, 0.10, 0.12, 0.14            & $10^2$ & $10^5$ \\
0.2 & $1792 \times 56$  & 0.16, 0.18, 0.2, 0.22, 0.24, 0.26 & $4 \times 10^2$ & $10^5$ \\
0.2 & $1792 \times 56$  & 0.28, 0.3, 0.32, 0.34, 0.36, 0.38 & $7 \times 10^3$ & $10^4$ \\
0.2 & $1792 \times 56$  & 0.4, 0.42, 0.44                   & $6 \times 10^4$ & $10^3$ \\
0.2 & $1792 \times 56$  & 0.48, 0.5, 0.52                   & $1.3 \times 10^5$ & $10^3$ \\
\hline
\end{tabular}
\caption{Summary of the simulation parameters.}
\label{table:simulation_parameter}
\end{table}

\begin{figure}[htb]
\centering
\includegraphics[width = 0.57\textwidth]{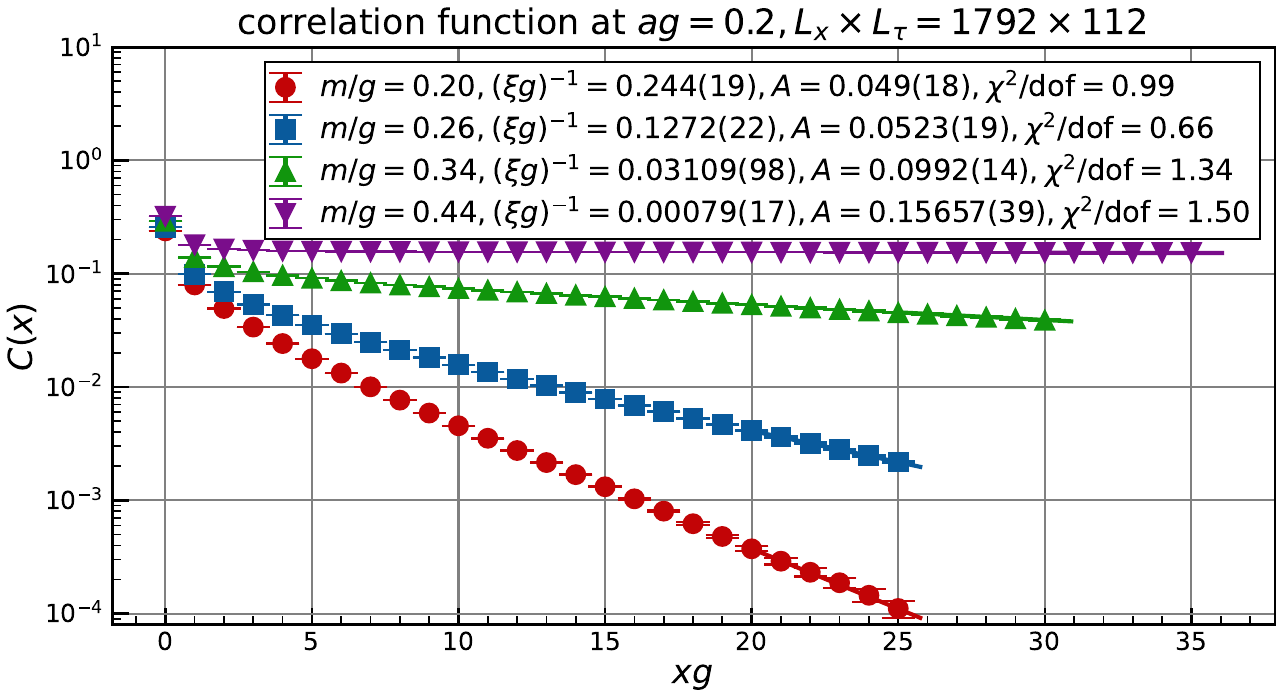}
\caption{
Correlation functions and fit results by the single exponential function~(\ref{eq:exponential}) at $m / g = 0.2, 0.26, 0.34, 0.44$. The lattice is $ag = 0.2, L_x \times L_\tau = 1792 \times 112$.
For visibility, only one-fifth of the data points are plotted.
\label{fig:fit_corr}
}
\end{figure}

For each correlation function, we perform a correlated fit using the single exponential function~(\ref{eq:exponential}) and extract the correlation length $\xi$ and amplitude $A$.
For demonstration, we show the correlation functions and their fit results at $m / g = 0.2, 0.26, 0.34, 0.44$ at the lower temperature in Fig~\ref{fig:fit_corr}.
The fit ranges are set as $xg \in \qty[20, 25], \qty[20, 25], \qty[25, 30], \qty[30, 35]$, for $m / g = 0.2, 0.26, 0.34, 0.44$, respectively.

\begin{figure}[htb]
\centering
\includegraphics[width = 0.45\textwidth]{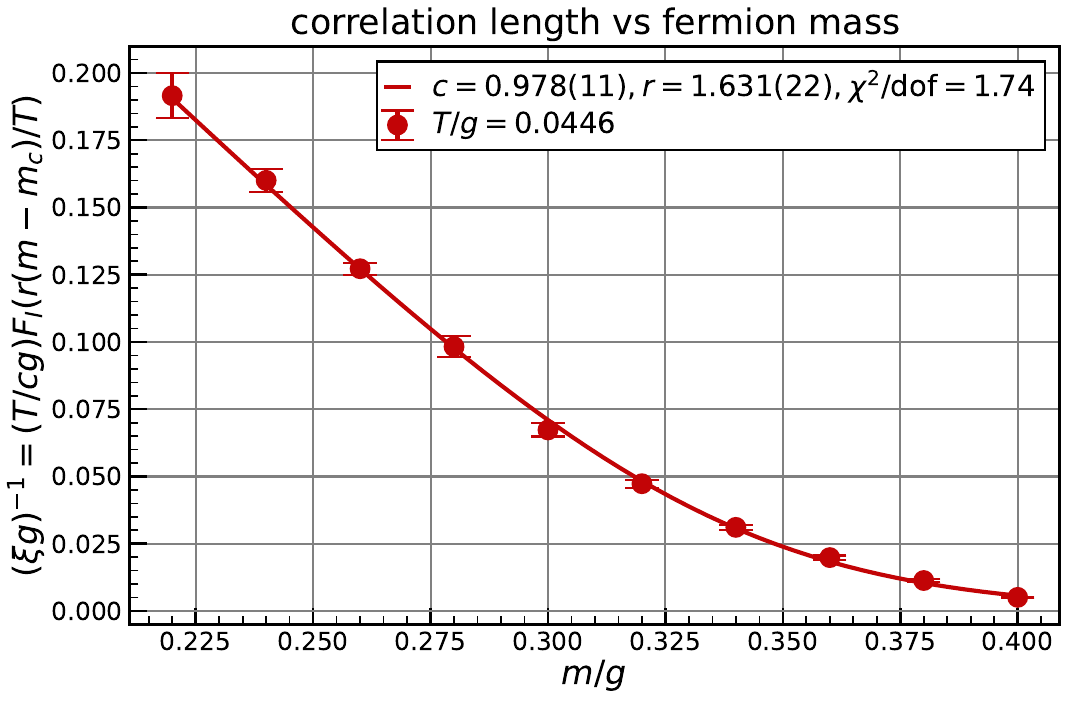}
\qquad
\includegraphics[width = 0.45\textwidth]{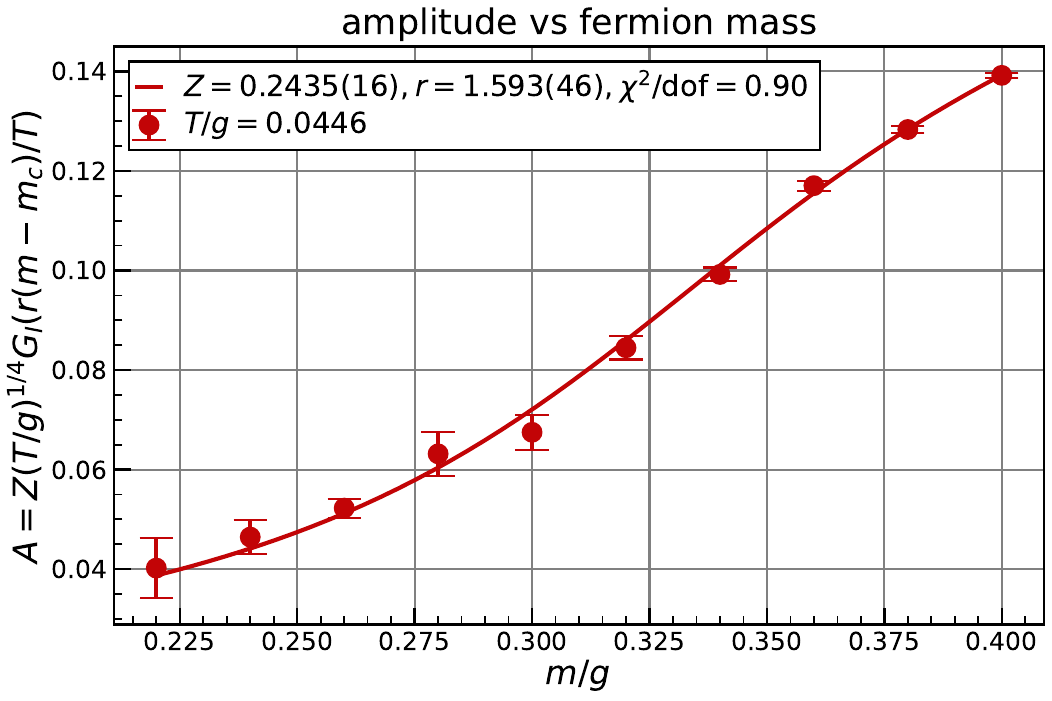}

\caption{
Fits to the correlation length (left) and to the amplitude (right) using the universal scaling functions of the quantum Ising chain~(\ref{eq:F}, \ref{eq:G}).
The lattice is $ag = 0.2, L_x \times L_\tau = 1792 \times 112$.
\label{fig:fit}
}
\end{figure}

We next perform fits to the correlation length and amplitude at the lower temperature, independently.
Universality with the quantum Ising chain suggests the asymptotic form of the correlation function in the Schwinger model at $\theta = \pi$ as
\begin{equation}
\lim_{x \to \infty} C(x) = Z \qty(T / g)^{1/4} G_I(\Delta / T) \exp(-\frac{Tx}{c} F_I(\Delta / T)), \quad \Delta = r (m - m_c), \label{eq:asymptotic_Schwinger}
\end{equation}
where $Z, c, r$ are nonuniversal dimensionless constants, and the functional forms of $F_I$ and $G_I$ are given in Eqs.~(\ref{eq:F}, \ref{eq:G}).
Figure~\ref{fig:fit} shows the correlation length and amplitude as functions of the fermion mass, fitted using 
\begin{equation}
\frac{T}{cg} F_I(r(m - m_c) / T), \quad Z(T / g)^{1/4} G_I(r(m - m_c) / T), \label{eq:fit_ansatz}
\end{equation}
respectively.
In these fits, we set the critical mass to $m_c / g = 0.3335$, using the wisdom obtained by Byrnes et al.~(\ref{eq:critical_mass}).
From the fits, the nonuniversal constants are found to be 
\begin{equation}
Z = 0.2435(16), \quad c = 0.978(11), \quad r = 1.593(46). \label{eq:nonuniversal}
\end{equation}
Note that the value of $r$ is also obtained from the fit to the correlation length, as shown in the left panel of Fig.~\ref{fig:fit}.
In the following analysis, however, we use the value obtained from the amplitude since the fit seems to be more reliable, although they are consistent with each other.

\begin{figure}[htb]
\centering
\includegraphics[width = 0.45\textwidth]{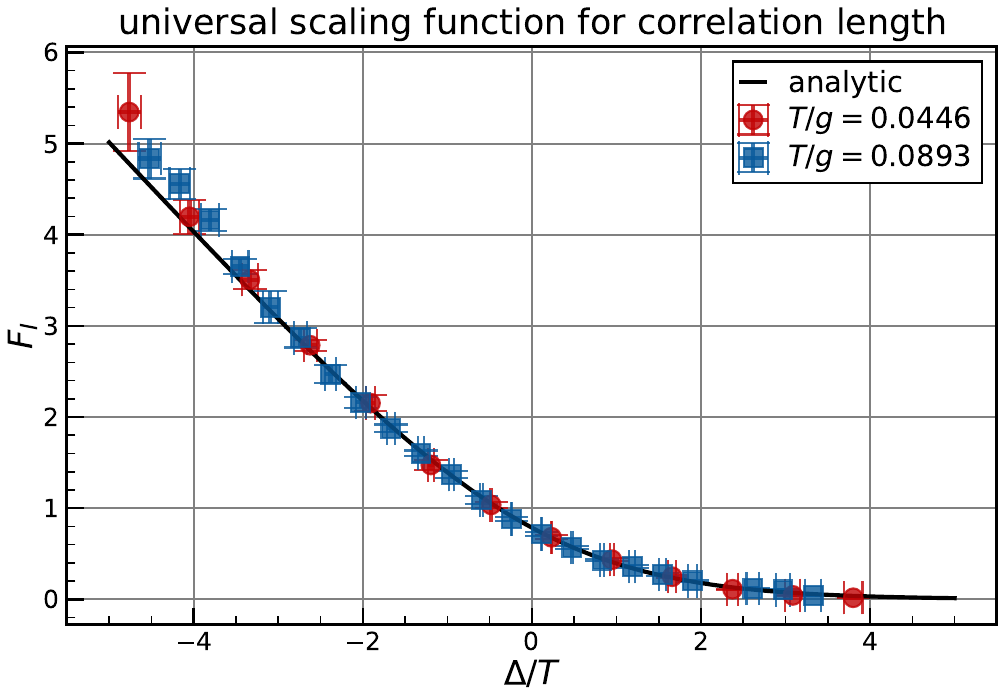}
\qquad
\includegraphics[width = 0.45\textwidth]{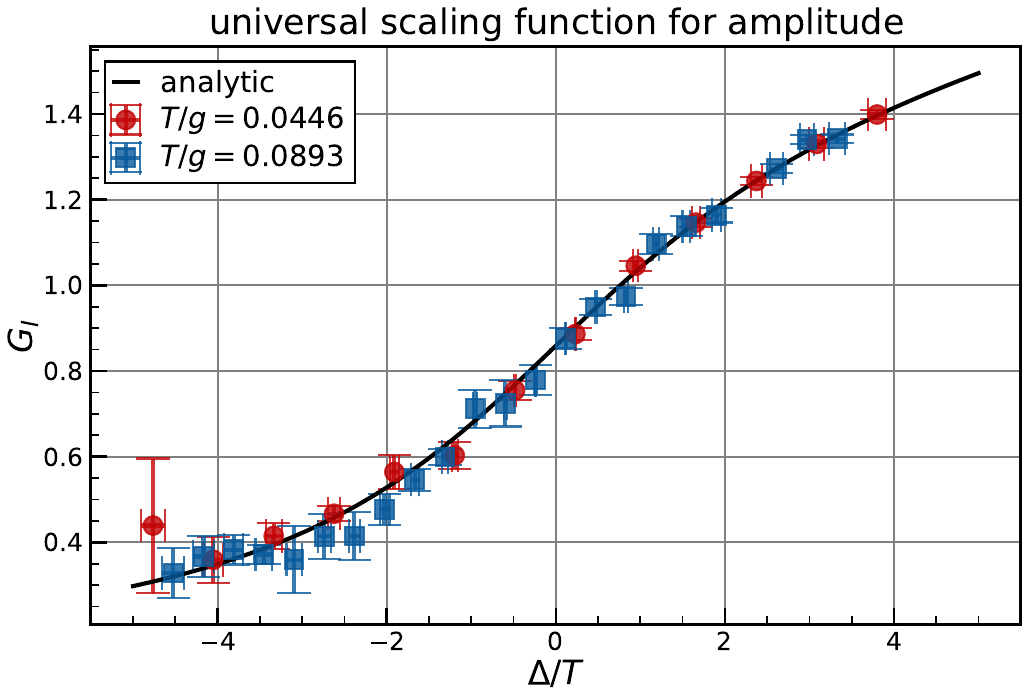}
\caption{
Rescaled correlation length (left) and amplitude (right) using the nonuniversal constants~(\ref{eq:nonuniversal}), 
compared with the universal scaling functions of the quantum Ising chain~(\ref{eq:F}, \ref{eq:G}).
\label{fig:universal_function}
}
\end{figure}

We rescale the correlation lengths and amplitudes at the two temperatures using the nonuniversal constants~(\ref{eq:nonuniversal}) and compare them with the universal scaling functions of the quantum Ising chain~(\ref{eq:F}, \ref{eq:G}). 
Figure~\ref{fig:universal_function} demonstrates that the rescaled data align beautifully with the expected analytical curves.
In particular, we observe that data at the higher temperature also match the analytical curves, 
providing strong evidence that the correlation function of the Schwinger model at $\theta = \pi$ shares the same asymptotic form as the quantum Ising chain.
Based on the argument in Section~\ref{sec:phase_diagram}, we conclude that the phase diagram of the Schwinger model at $\theta = \pi$ is entirely analogous to the quantum Ising chain near the QCP,
as schematically shown in Fig.~\ref{fig:phase_diagram}.

\subsection{Range of scaling behavior}
With the established asymptotic form of the correlation function~(\ref{eq:asymptotic_Schwinger}) and the explicit values of the nonuniversal constants~(\ref{eq:nonuniversal}), 
we can quantitatively predict the behavior of certain observables near the QCP.
In this subsection, we roughly estimate the range of the scaling behavior by comparing the predictions to direct numerical results.

We first examine how well the scaling behavior holds at zero temperature by comparing our prediction with Byrnes et al.'s results~\cite{Byrnes:2002nv}.
\footnote{The explicit values can be found in Table~6.10 of Ref.~\cite{Byrnes:phd}.}

\begin{figure}[htb]
\centering
\includegraphics[width = 0.45\textwidth]{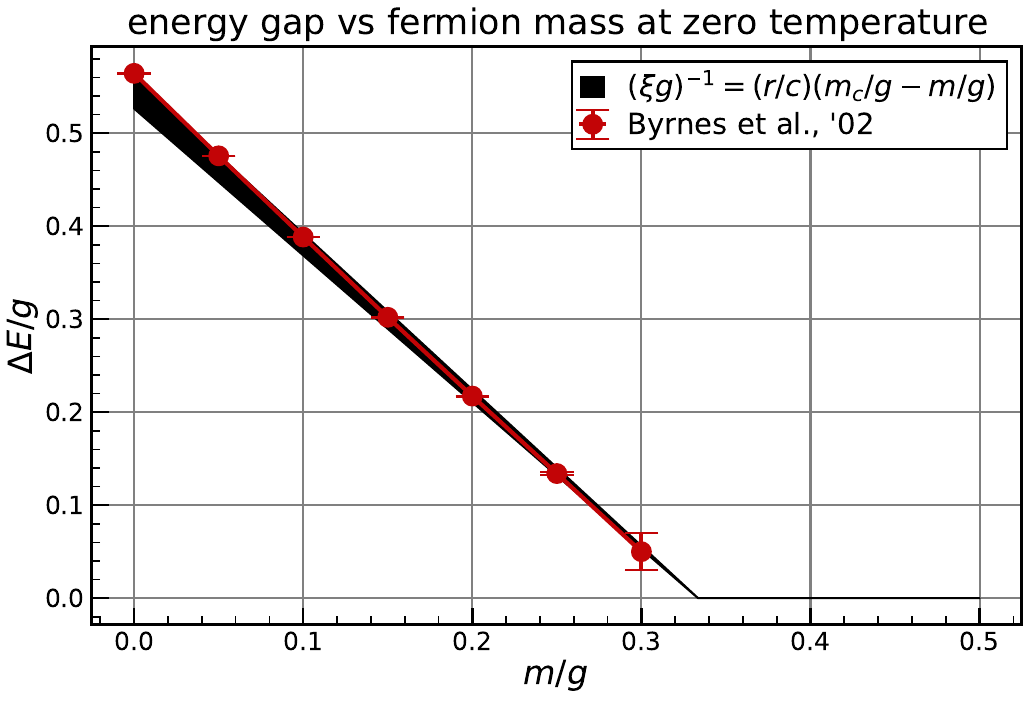}
\caption{
Fermion mass dependence of the energy gap at zero temperature.
\label{fig:energyGap}
}
\end{figure}

At zero temperature in the CP symmetric phase ($m < m_c$), the inverse of the correlation length, i.e., the energy gap, is given by
\begin{equation}
\qty(\xi g)^{-1} = \frac{1}{c} \frac{\abs{\Delta}}{g} = \frac{r}{c} \qty(\frac{m_c}{g} - \frac{m}{g}). \label{eq:Ohata}
\end{equation}
In Fig.~\ref{fig:energyGap}, we compare Eq.~(\ref{eq:Ohata}) with their direct numerical results~\cite{Byrnes:2002nv}.
Figure~\ref{fig:energyGap} demonstrates a strong agreement between the prediction and their direct numerical results, 
indicating that the scaling behavior holds surprisingly well in the region $m \in \qty[0, m_c]$ at zero temperature.

\begin{figure}[htb]
\centering
\includegraphics[width = 0.45\textwidth]{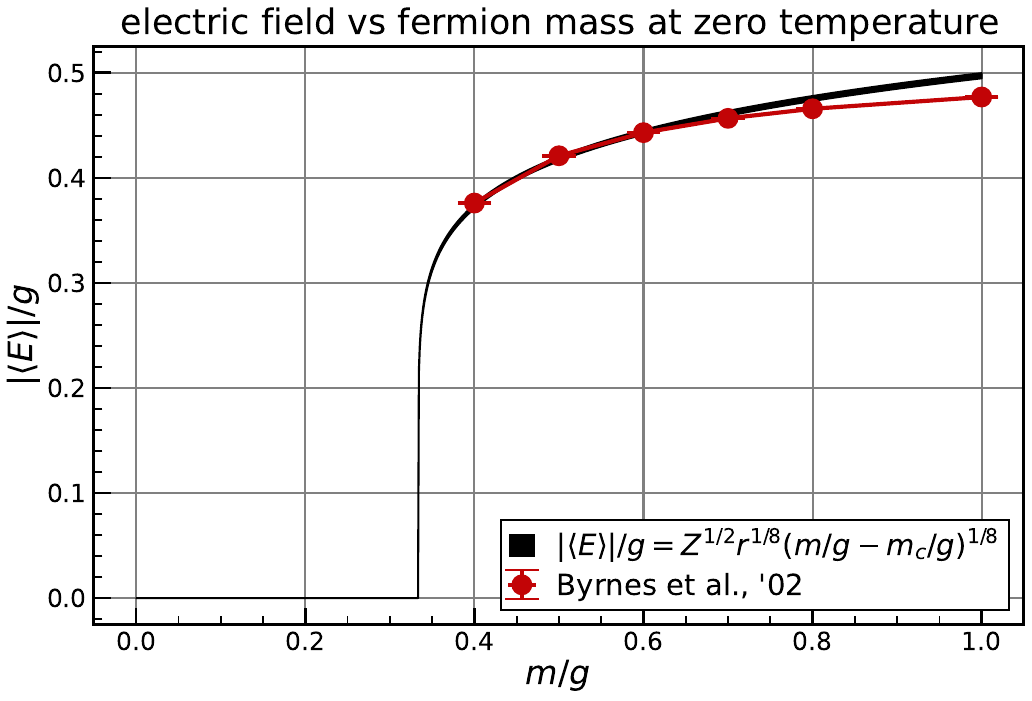}
\caption{
Fermion mass dependence of the electric field at zero temperature.
\label{fig:order_parameter}
}
\end{figure}

We can also predict the electric field in the CP broken phase ($m > m_c$) at zero temperature from the asymptotic form of the correlation function~(\ref{eq:asymptotic_Schwinger}).
In this region, using the zero temperature limit of $G_I$~\cite{Sachdev:1995bf}
\begin{equation}
G_I(\Delta / T) \to (\Delta / T)^{1/4}, \quad \Delta / T \gg 1,
\end{equation}
the correlation function behaves as
\begin{equation}
C(x) \to Z(T / g)^{1/4} \qty(\Delta / T)^{1/4} = Z \qty(\Delta / g)^{1/4} = Z r^{1/4} \qty(\frac{m}{g} - \frac{m_c}{g})^{1/4}.
\end{equation}
Therefore, the electric field in the CP broken phase is given by
\begin{equation}
\frac{\abs{\expval{E}}}{g} = Z^{1/2} r^{1/8} \qty(\frac{m}{g} - \frac{m_c}{g})^{1/8}. 
\label{eq:order_parameter}
\end{equation}
We plot Eq.~(\ref{eq:order_parameter}) using the nonuniversal constants~(\ref{eq:nonuniversal}) in Fig.~\ref{fig:order_parameter}.
We observe a rather sharp but continuous increase in the electric field near the QCP, 
corresponding to the small critical exponent $\beta = 1/8$.
In Fig.~\ref{fig:order_parameter}, the numerical results by Byrnes et al.~\cite{Byrnes:2002nv} are also plotted. We find that the analytical curve successfully explains their numerical results near the QCP.
While clear discrepancies are observed away from the QCP, the discrepancies are very small even at $m / g = 1$. 
This allows us to conclude that the scaling behavior holds well in the region $m / g \in \qty[m_c / g, 1]$ at zero temperature.

The agreements with Byrnes et al.'s results near the QCP reinforce the validity of our estimate for the nonuniversal constants~(\ref{eq:nonuniversal}) and indicate that the scaling behavior holds well in the region $m / g \in [0, 1]$ at zero temperature.

\begin{figure}[htb]
\centering
\includegraphics[width = 0.45\textwidth]{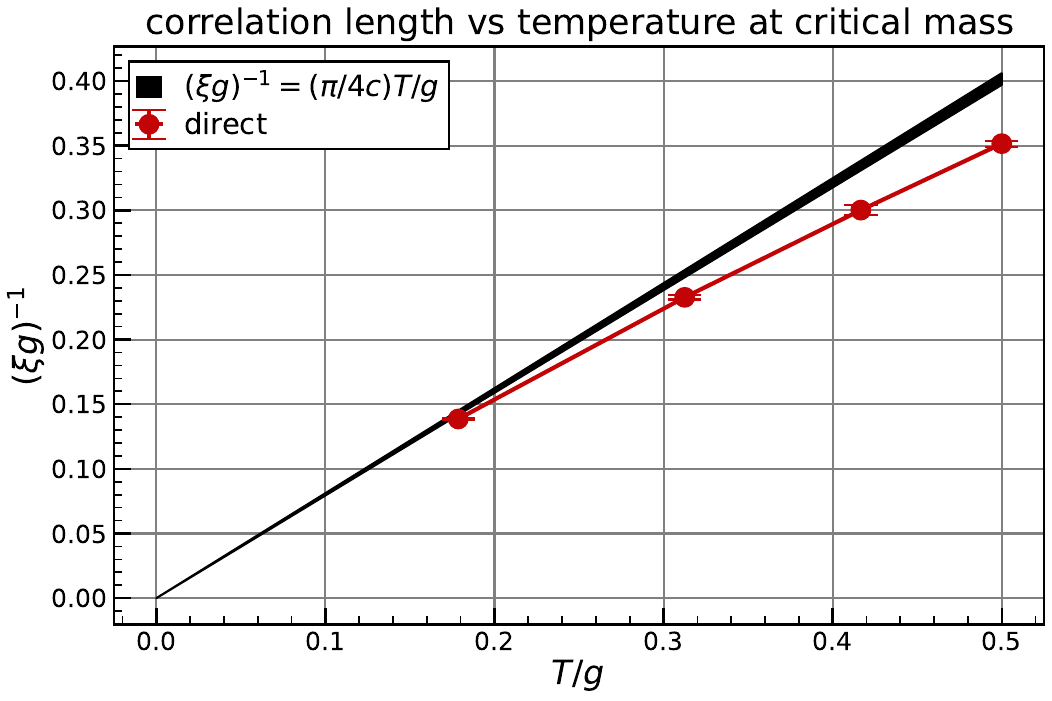}
\qquad
\includegraphics[width = 0.45\textwidth]{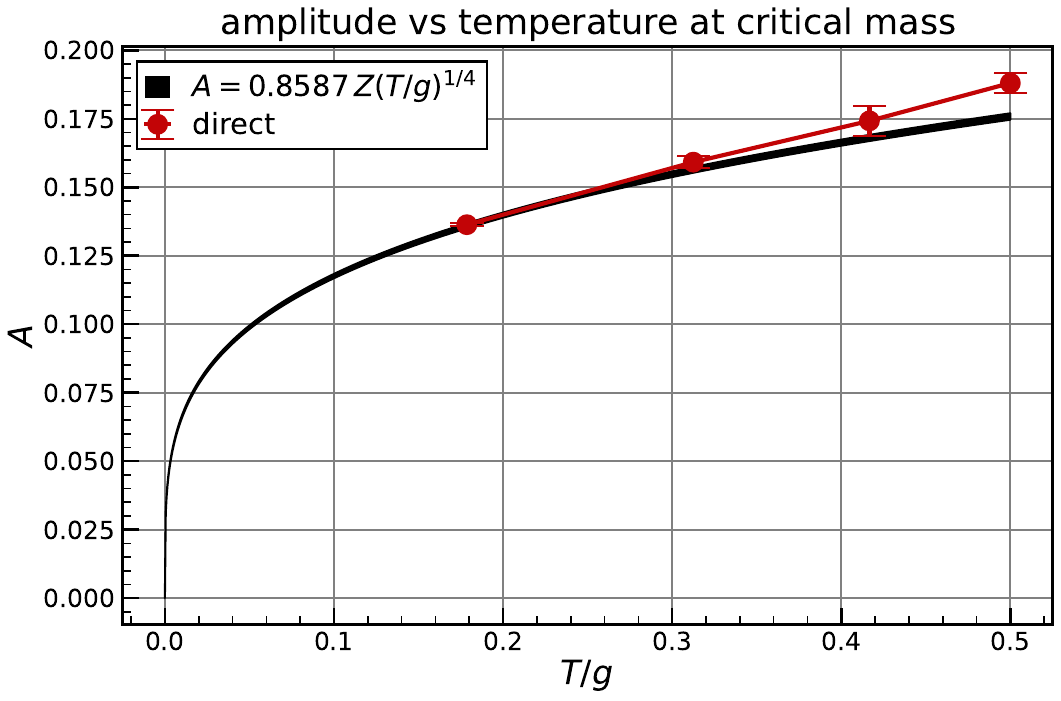}

\caption{
Temperature dependence of the correlation length (left) and the amplitude (right) at the critical fermion mass.
\label{fig:temperature_dependence}
}
\end{figure}

We also examine the range of the scaling behavior at the critical fermion mass.
In this region, the inverse of the correlation length is
\begin{equation}
\qty(\xi g)^{-1} = \frac{\pi T}{4c g}, \label{eq:temp_corre}
\end{equation}
and the amplitude is given by
\begin{equation}
A = 0.8587 Z \qty(\frac{T}{g})^{1/4}, \label{eq:temp_amp}
\end{equation}
from the behavior of $G_I$~\cite{Sachdev:1995bf}:
\begin{equation}
G_I (\Delta / T) = 0.8587..., \quad \abs{\Delta} / T \ll 1.
\end{equation}
In Fig.~\ref{fig:temperature_dependence}, we compare Eqs.~(\ref{eq:temp_corre}, \ref{eq:temp_amp}) with our direct numerical results, which are obtained in the same way as in the previous subsection.
The simulation parameters are summarized in Table.~\ref{table:simulation_parameter_temperature}.
\begin{table}[!h]
\centering
\begin{tabular}{cccccc}
\hline
$ag$ & $L_x \times L_t$ & $m / g $ & $\tau_{\mathrm{int}}$ & num. of iterations & num. of configurations \\
\hline
0.2 & $1792 \times 28$ & 0.3335 & $\sim 2 \times 10^3$ & $10^3$ & $10^5$ \\
0.2 & $1792 \times 16$ & 0.3335 & $\sim 10^3$ & $5 \times 10^2$ & $10^5$ \\
0.2 & $1792 \times 12$ & 0.3335 & $\sim 5 \times 10^2$ & $2.5 \times 10^2$ & $10^5$ \\
0.2 & $1792 \times 10$ & 0.3335 & $\sim 3 \times 10^2$ & $1.5 \times 10^2$ & $10^6$ \\
\hline
\end{tabular}
\caption{Simulation parameters and estimated autocorrelation times for investigating temperature dependence of the correlation length and the amplitude at the critical fermion mass.}
\label{table:simulation_parameter_temperature}
\end{table}
We find that the direct numerical results agree with the analytical curves with an accuracy of at least ninety percent up to $T / g \simeq 0.4$.
This ensures that the scaling behavior holds well in the region $T / g \in [0, 0.4]$ at the critical fermion mass.

\begin{figure}[htb]
\centering
\includegraphics[width = 0.45\textwidth]{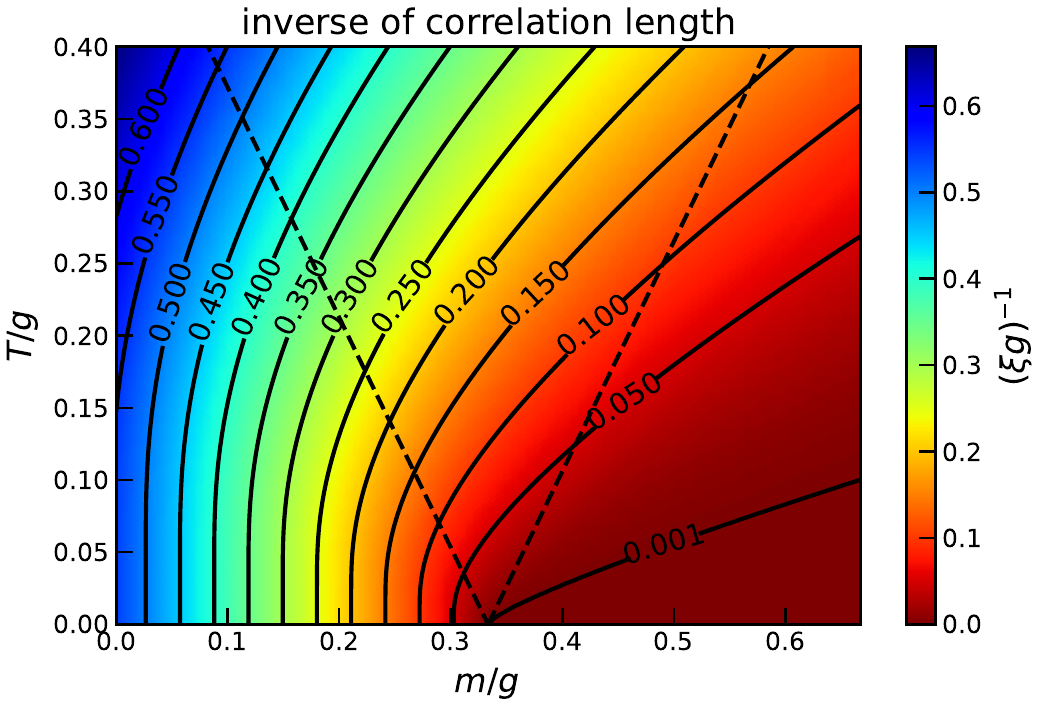}
\qquad
\includegraphics[width = 0.45\textwidth]{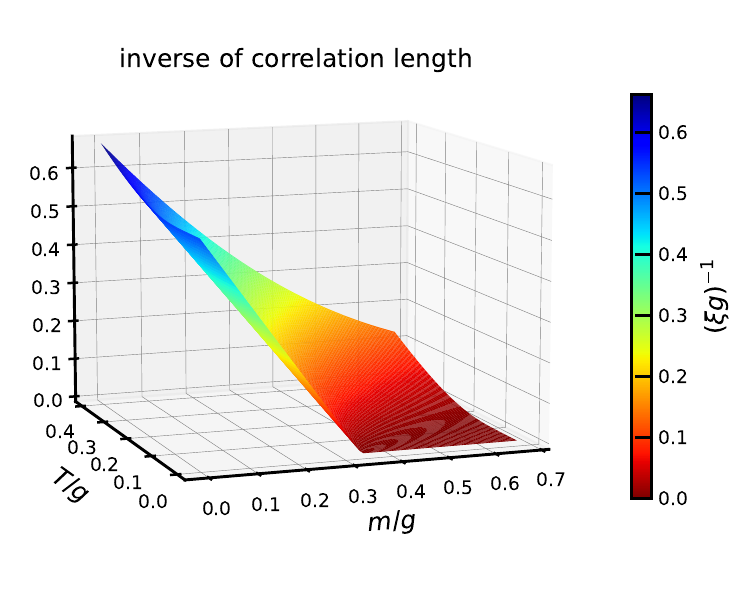}

\caption{
Inverse of the correlation length in the temperature and fermion mass plane obtained from the asymptotic form~(\ref{eq:asymptotic_Schwinger}) and the nonuniversal constants~(\ref{eq:nonuniversal}).
The dashed lines $T = \pm\Delta, \, \Delta = r(m - m_c)$ in the left panel are the crossover lines that classify the phases effectively.
\label{fig:3d_plot}
}
\end{figure}

In Fig.~\ref{fig:3d_plot}, we finally present the inverse of the correlation length in the temperature and fermion mass plane obtained from the asymptotic form~(\ref{eq:asymptotic_Schwinger}) and the nonuniversal constants~(\ref{eq:nonuniversal}).
Based on the analysis in this subsection,
the values are reliable at zero temperature or at the critical fermion mass within the range plotted in Fig.~\ref{fig:3d_plot}.
At $m = 0$, the Schwinger model is equivalent to the free boson theory~(\ref{eq:continuum_hamiltonian}), resulting in an energy gap of $g / \sqrt{\pi} \simeq 0.564 g$ at any temperature.
In Fig.~\ref{fig:3d_plot}, we indeed observe approximate temperature independence up to $T / g \simeq 0.3$, suggesting that Fig.~\ref{fig:3d_plot} is to some extent reliable except for the two edges on the finite-temperature side.


\section{Summary and conclusion} \label{sec:summary}
In this paper, we have explored the phase diagram of the Schwinger model at $\theta = \pi$ in the temperature and fermion mass plane.
We first reviewed the spontaneous CP symmetry breaking at zero temperature in the Schwinger model at $\theta = \pi$ and revisited the phase diagram of the quantum Ising chain in the temperature and coupling constant plane.
There, we observed that the phase diagram of the quantum Ising chain can be deduced from the explicit asymptotic form of the correlation function~(\ref{eq:asymptotic_Ising}).
From the perspective of universality, we then conjectured the phase diagram of the Schwinger model at $\theta = \pi$ near the QCP, which is entirely analogous to the quantum Ising chain.
This conjecture was subsequently confirmed by validating that the correlation function of the electric field shares the same asymptotic form as the quantum Ising chain through first-principle Monte Carlo simulations.
We circumvented the sign problem, which hinders the conventional Monte Carlo simulation at $\theta \neq 0$, by using the lattice formulation of the bosonized Schwinger model.
Furthermore, we examined the range of the scaling behavior by comparing the predictions based on universality to direct numerical results and identified a surprisingly wide applicable region.

Our present study has demonstrated the power of universality in a gauge theory.
In fact, the data used to determine the nonuniversal constants~(\ref{eq:nonuniversal}) were obtained through Monte Carlo simulations at a single finite temperature.
Nevertheless, through universality with the quantum Ising chain, we were able to quantitatively predict the critical behavior of the electric field at zero temperature, as shown in Fig.~\ref{fig:order_parameter}.
This prediction closely matched the direct numerical results obtained from a completely different formulation~\cite{Byrnes:2002nv} near the QCP.
We hope that our work serves as a successful application example of universality in a gauge theory
and inspires similar research for understanding critical behaviors, not only in the Schwinger model but also in other gauge theories, including quantum chromodynamics.

\ack
The author was supported by a Grant-in-Aid 
for JSPS Fellows (Grant No.22KJ1662).
The numerical simulations have been carried out on Yukawa-21 at Yukawa Institute for Theoretical Physics, Kyoto University.

\bibliography{PhaseDiagram.bib}
\end{document}